# Listen to Users, but Only 85% of the Time: How Black Swans Can Save Innovation in a Data-Driven World

**Maximilian Speicher**
Jagow Speicher, Barcelona, Spain
arxiv@maxspeicher.com

**Abstract**

Data-driven design is a proven success factor that more and more digital businesses embrace. At the same time, academics and practitioners alike warn that when virtually everything must be tested and proven with numbers, that can stifle creativity and innovation. This article argues that Taleb's Black Swan theory can solve this dilemma. It shows that online experimentation, and therefore digital design, are fat-tailed phenomena and, hence, prone to Black Swans. It introduces the notion of *Black Swan designs*—"crazy" designs that make sense only in hindsight—along with four specific criteria. To ensure incremental improvements and their potential for innovation, businesses should apply Taleb's barbell strategy: Invest 85–90% of resources into data-driven approaches and 10–15% into potential Black Swans.

**Original Publication**

Over the past two decades, we have witnessed a shift from classical design to what John Maeda calls computational design (Maeda, 2019: pp. xi-xii). With this shift—and amplified by works such as *The Lean Startup* (Ries, 2008) and *Sprint* (Knapp, Zeratsky, & Kowitz, 2016)—came a strong focus on minimum viable products (MVPs), continuous iteration, and user testing. One specifically popular form of testing in this context is online experimentation (also called A/B testing). That is, an (un-)finished digital design can be put live first and then quantitatively evaluated with huge amounts of real users. The possibility to do this is one of the core features of computational design (Maeda, 2019: pp. 155-157). Businesses can benefit significantly from leveraging this approach. For instance, Sheppard, Sarrazin, Kouyoumjian, & Dore (2018) have found that companies that employ data-driven design, continuous iteration, and user centricity "[outperform] industry-benchmark growth by as much as two to one." All of these have become staples in many digital businesses that live by the rule of "if it's not tested it's just another opinion." In this respect, especially online experimentation is sometimes considered a jack-of-all-trades solution (Speicher, 2021a). There are companies today that have such extensive experimentation infrastructures that one can assume virtually everything is an online experiment (cf. Bakshy, Eckles, & Bernstein, 2014; Kohavi et al., 2013; Xu, Chen, Fernandez, Sinno, & Bhasin, 2015).

However, Don Norman as well as Saul Greenberg and Bill Buxton recognized as early as 2005 and 2008, respectively, that there can indeed be such a thing as too much focus on users and data, which hinders innovation (Norman, 2005; Greenberg & Buxton, 2008). This is reflected by more recent writings from industry practitioners reporting that the ubiquity of testing and validation and other data-driven design practices stifle creativity and innovative ideas. To give just three examples, Sun (2022) observes that "[t]he rise of data-driven culture cultivated a generation of designers who only take risk-free and success-guaranteed steps towards the inevitable local maxima of design monotony." Likewise, Buckley (2022b) reports that in contemporary industry settings, when everything is validated with users, not enough risks are taken and, therefore, "[m]odern UX design is killing creativity." Lisefski (2019) states the same observation differently: "We're told all design decisions must be validated by user feedback or business success metrics. [...] If it can't be proven to work in a prototype, A/B test, or MVP, it's not worth trying at all."

Upon closer inspection, both sides have a point. The shift to computational design and data-driven organizations—merging in data-driven design—has yielded enormous value for businesses in the digital age (Sheppard et al., 2018). However, it can also not be in their best interest to lose innovation, creativity, and "designer instinct" (Lisefski, 2019) to excessive data-drivenness and therefore miss out on potentially groundbreaking ideas that could disrupt industries.

What could be a solution to this dilemma? One possibility is: *Listen to users, but only 85% of the time.* That is, invest a majority of resources in the kind of design that has made today's digital businesses so successful. But reserve the remainder for *testing a specific, new kind of "crazy" designs* that can bring about radical innovation. This stems directly from applying the theory of Black Swans (Taleb, 2007) to digital design, as I will detail in this article.

More specifically, I will present thoughts on

1. why outcomes of online experiments follow a fat-tailed distribution and how, therefore, Nassim Nicholas Taleb's theory of the Black Swan (Taleb, 2007) can be applied to digital design;
2. four specific criteria digital designs have to fulfill to be able to have a huge impact and potentially bring about radical, or even disruptive, innovation; and
3. how such *Black Swan designs* and corresponding *Black Swan experiments* could be embedded into a barbell strategy (Taleb, 2007: ch. XIII) so as to not hurt sensitive business metrics and still bear the potential to innovate.

In the following, I will first provide an outline of online experimentation and data-driven design and the reasons why a data-driven approach to design can hinder innovation. Subsequently, I will introduce the concept of the Black Swan. In the third section, I will show that online experiments follow a fat-tailed distribution, apply Black Swan theory to digital design, and deduce criteria for characterizing Black Swan designs. The fourth section explains how such designs and corresponding Black Swan experiments can be embedded into a more conservative framework to balance incremental improvements with a potential for innovation. Finally, I address related work and give concluding remarks.

# 1. Online Experimentation & Data-Driven Design

Online experimentation is at the core of how many businesses interpret data-driven design. Often, the two are inevitably entangled. In the following, I will elaborate on why one cannot talk about data-driven design without also talking about online experimentation and why current practices are problematic from an innovation standpoint.

## 1.1. Online Experimentation

In an online experiment, or A/B test, different variations (A, B, ..., n) of a digital product are randomly distributed among users and measured against one or more KPIs, such as conversions or newsletter subscriptions. A very simple A/B test could be to show 50% of users a green "subscribe" button while the other 50% see a red button, with the underlying hypothesis that the green button will lead to (statistically) significantly more subscription events due to the more positive coloring that communicates "OK" rather than "error."

While online experiments can be used to test changes to any aspect of a digital product—including those entirely invisible to the user—in this article, I only refer to online experiments that test (changes to) a digital design. This is, anything that a digital design team can influence in terms of information architecture, interaction design, visual design, etc.

## 1.2. Data-Driven Design

When I say "data-driven design," I more or less refer to what Lisefski (2019) describes: "[...] all design decisions must be validated by user feedback or business success metrics. [...] If it can't be proven to work in a prototype, A/B test, or MVP, it's not worth trying at all." Analogously, one CEO I once worked with said: "If it hasn't proven its value in a test, it doesn't have value." And indeed, if companies follow the criteria laid out by Sheppard et al. (2018),[1] every potentially significant design or change to a digital product must be tested and measured at some point. Online experimentation is a favorite for such measurements due to the relative ease with which experiments can be deployed. Additionally, the fact that one can collect large amounts of quantitative data with real users appeals to businesses. The triumphal march of online experimentation has even spawned an entirely new industry around "conversion rate optimization" services (cf. Speicher, 2021a). But quickly evaluating concepts, screen designs, and prototypes of different fidelities has also become much easier due to the proliferation of online platforms for remote user research.

---

[1] "Measure and drive design performance with the same rigor as revenues and costs" and "De-risk development by continually listening, testing, and iterating with end-users," among others (Sheppard et al., 2018).

Data-driven design can take on many forms. Some data-driven businesses use mainly remote usability testing, some do mainly online experiments, and many a mix of these two. According to Speicher (2021b), companies falling into the second category practice *KPI-centered design*. This methodology is characterized by an almost exclusive focus on business-critical metrics (so-called key performance indicators, or KPIs), hypotheses based on assumptions, and online experiments as the primary means of user research for testing those hypotheses (Speicher, 2021b).

Today, one particularly popular form of data-driven design that is often advertised by businesses specializing in online experimentation services (e.g., Patel, 2021; VWO, 2022) is *evidence-centered design* (Speicher, 2021b). That is, rather than being based on "wild" assumptions, hypotheses are backed by evidence—a best practice, a competitive analysis, customer insights, an earlier user study, etc. Additionally, online experimentation is complemented with additional, usually more qualitative, user research. This is a step forward from purely KPI-centered design when it comes to bringing the user into the equation. However, it is a very conservative approach in the sense that it reduces the inherent riskiness of online experimentation. For instance, Stroud (2021) states that "when some form of UX research with real users went into building the hypothesis or contributing to the designs being A/B tested," one can considerably increase the rate of tests with a positive result. That, in turn, means making experimentation more predictable, which can further reduce the potential for innovation.

## 1.3. Why Can Too Much Data-Drivenness Work Against Innovation?

All of this is in line with the observations of Buckley (2021), Lisefski (2019), and Sun (2022) as well as the concerns of Norman (2005) and Greenberg and Buxton (2008). The latter write:

> "... evaluation can be ineffective and even harmful if naively done 'by rule' rather than 'by thought'. If done during early stage design, it can mute creative ideas that do not conform to current interface norms. If done to test radical innovations, the many interface issues that would likely arise from an immature technology can quash what could have been an inspired vision." (Greenberg & Buxton, 2008)

But why can too much focus on data and users be harmful? There are three factors at play.

First, users are, in general, bad at recognizing and describing what they really want and need (Norman, 2005). Additionally, they often have an inherent "innovation resistance" that is a result of status quo bias (Kahneman, Knetsch, & Thaler, 1991; Stryja, Dorner, & Riefle, 2017). That is, users have to get accustomed to innovative designs and solutions over a longer period of time before they can appreciate them. For instance, runtimes of 14 days to one month for online experiments, which are the usual case in industry, are most probably not suitable for testing larger design changes that might be subject to innovation resistance, as I have experienced first-hand.[2]

Second, many user tests and validations that are carried out are "lean" or "quick 'n' dirty." This is often a result of a general shortage of resources or of practices like design sprints (cf. Knapp et al., 2016) that have very short iteration cycles with tight deadlines. Unfortunately, this often inevitably leads to methodological issues with the experimental design or participant recruitment, which prevent researchers from obtaining valid results and drawing the correct conclusions.

Third, the business KPIs that are employed in data-driven design—and especially in online experimentation—are very often short-term in nature (e.g., click-through rate, conversion rate, or revenue). Therefore, they are not able to capture longer-term effects of innovative solutions. For instance, focusing too much on conversion rates can worsen the user experience (Speicher, 2021a).

---

[2] I recall numerous online experiments I have observed that replaced a design with obvious flaws with an improved one based on common design principles and heuristics and still, results were negative after runtimes of 2–4 weeks. One specific example is the replacement of a website navigation with significant performance issues with a faster, state-of-the-art burger menu. Even after four weeks, metrics suggested that, on average, users had not yet become familiar with the new navigation.

`booking.com` and `ryanair.com` are prime examples of this, with numerous dark patterns (cf. Narayanan, Mathur, Chetty, & Kshirsagar, 2020) that have most probably been very thoroughly tested but make for a frustrating experience. In general, an unhealthy fixation on KPIs can hurt users and the business alike (Laubheimer & Moran, 2021).

All of these factors favor incremental, predictable improvements over bigger, more innovative solutions. Padua (2018) states this as follows:

> "'Prove it' takes innovation off the table because the business is effectively looking for a predictable outcome; it is looking not for innovation but for an incremental step along the existing business path, a reinforcement of the 'what has made this business successful to date, will continue to work'. Not only is that no longer true in today's disrupted economy; it is also most certainly not innovation. [...] Experimentation has no certainty of success, no known expected outcome."

Imagine a design that is an entirely novel solution to a significant problem of a specific user group. Because the design resembles nothing anyone has ever seen, that user group would need a settling-in period of three months, and only after that period would a business see a noticeable impact. How should a remote usability test of an early prototype with 100 participants, 50 of which do not belong to the target group, be able to capture the potential of the situation? Or an MVP that is deployed by means of an online experiment with a runtime of one month? This also illustrates that data-drivenness *per se* is not necessarily the problem, but rather the reliance on a relatively narrow set of ill-suited user research setups.

When every online experiment must be backed by evidence already before it is conducted and every design must be validated with users before it is considered for implementation, that eliminates the inherent uncertainty and unpredictability that can bring about innovative changes. In the following, I will investigate how Black Swans can help us mitigate the above shortcomings, and how an approach might look like that combines the best of both worlds—data-drivenness and the potential for (radical) innovation.

## 2. What is a Black Swan?

To give a very brief explanation, according to Taleb, a Black Swan is a "highly improbable" event that fulfills three properties (Taleb, 2007: pp. 19-20):

1. "it lies outside the realm of regular expectations, because nothing in the past can convincingly point to its possibility"
2. "it carries an extreme 'impact'"
3. Humans find it "explainable and predictable" in retrospect due to hindsight bias (cf. Roese & Vohs, 2012).

In relation to this, Taleb (2007: pp. 70-84) also talks about "Mediocristan" and "Extremistan." While in the former, phenomena usually follow a normal distribution and are, therefore, somewhat predictable (e.g., a human's body height), in the latter, extreme outliers skew the distribution so much that it is not meaningful anymore to talk about averages (e.g., distribution of wealth). According to the three properties above, Black Swans can only occur in Extremistan. Notably, they can be negative (the Black Monday) as well as positive (the unicorn startup no-one believed in).

One very specific example of a Black Swan event is the rise of personal computing (Strategic CFO™, 2015). In *What the Dormouse Said*, Markoff (2005) details how it was anything but inevitable. Many coincidences had to happen, more or less at the same time and in the same place, for this revolution to be set in motion, against all odds, including pushbacks from major corporations. I believe that I do not need to discuss the impact personal computing has had on the world and that, in hindsight, it seems relatively obvious to invent such a thing as a personal computer. Hence, all three of Taleb's criteria are fulfilled.

## 3. What Can Digital Design Learn from Black Swans?

The kind of online experiments addressed in this article test (changes to) digital designs. This means that they quantify the impact a digital design carries (cf. Taleb's criterion no. 2). Hence, to be able to apply Black Swan theory to digital design, we need to show that such online experiments are a phenomenon of Extremistan. That is, the effect sizes of a set of online experiments must adhere to a fat-tailed distribution (cf. Taleb, 2020). Here, "effect size" means the result of a single experiment with respect to the defined primary metric. For instance, if the control version has an observed conversion rate of 5% and the treatment has an observed conversion rate of 5.5%, the effect is an uplift with an effect size of 10%.

## 3.1. Digital Designers Live in Extremistan

As per my experience of having conducted and/or observed hundreds of online experiments, we are certainly *not* talking about Mediocristan here. There is no reason to assume that effect sizes of online experiments are normally distributed. First, websites have very different levels of design maturity that change over time. The higher your room for improvement, the easier it is to find "low-hanging fruit" experiments that produce statistically significant uplifts. Second, experiments can vary greatly in the changes they are introducing: some just change the color of a button, some add a whole new feature. Lastly, different experiments focus on different primary metrics: some aim at increasing revenue, some at newsletter subscriptions, and some at clicks on a banner. You can conduct a series of "best practice" tests from any given online experimentation playbook[3] and the results will probably look kind of normally distributed and lie mostly between ±0% and +5%; but anytime, one radical experiment could, against all odds, produce an uplift of 200% (Kohavi, Deng, Longbotham, & Xu, 2014). We can illustrate this by looking at an exemplifying meta-analysis of 115 publicly available experiments (Georgiev, 2018). They show effect sizes that somewhat resemble a normal distribution, but with a long right tail (Georgiev, 2018). Azevedo, Deng, Montiel Olea, Rao, and Weyl (2020) produced corroborating results based on an analysis of 1,450 experiments. Unlike for, e.g., human body height, there is theoretically no limit on how large the effect of an online experiment could become.

Hence, if you are a digital designer, you live in Extremistan. Black Swans happen in online experimentation, and therefore digital design, and we should leverage that for truly impactful innovation.

---

[3] For instance, the Dynamic Yield Inspiration Library (https://www.dynamicyield.com/personalization-examples/, accessed June 27, 2022) or GoodUI Patterns (https://goodui.org/patterns/, accessed June 27, 2022).

## 3.2. Black Swan Designs

Let us characterize such digital designs that bring about positive Black Swans in analogy to Taleb (2007: pp. 19-20).

**Black Swan designs** are digital designs that

1. are not based on evidence and do not appear in best practices, guides, competitive analyses, inspiration libraries, or playbooks of any sort;
2. have an extreme impact; and
3. virtually no-one believes in beforehand (but afterwards they say: "I knew this would work all along").

A Black Swan design does not necessarily have to introduce big changes, since even small ones can have a huge impact (Kohavi et al., 2014). Moreover, *the fewer people would bet on a design* to actually work (those ideas that are put at the very bottom of the backlog in a refinement because nobody believes in them), the more likely it is a candidate for a Black Swan. However, while being radical, such a design also has to be functional, which is an additional fourth criterion. For instance, black text on a black background would be radical but render any website unfunctional.

Hence, why not try a reverse checkout for e-commerce customers? There is no evidence whatsoever indicating users might want that; and I hypothesize that virtually nobody would dare to ask for the resources to test it. Or maybe "rush-hour shopping," with 5 random products at a time and a "next" button, and you get 20% off everything you add to your basket within 3 minutes? It might work for your users and/or brand in ways you cannot imagine. I must admit here that I find it extremely difficult to come up with examples of Black Swan designs—and the two I have given are probably not good ones. The reason is that it is safe to assume one is always biased towards things one subconsciously assumes to work when ideating. This, however, also highlights one of the challenges of radical or disruptive innovation. Often, the best ideas can only be identified after the fact rather than planned. Better examples for Black Swan designs could probably be found in the graveyard of design ideas that were deleted from backlogs in my current and former jobs (however, I cannot disclose those for reasons of confidentiality).

How it might be possible to consistently identify potential Black Swan designs—i.e., how to operationalize the theory laid out in this article and make those Black Swans gray (Taleb, 2007: pp. 396-397)—is subject to further research. I hope to motivate such research with this article.

### 3.3. Black Swan Experiments

It is vital for potential Black Swan designs to be tied to a corresponding **Black Swan experiment**. Such experiments must be longitudinal in order to overcome potential innovation resistance. Otherwise, we have no means of determining whether a design is indeed a (positive) Black Swan design or not. Because of the inherent riskiness and uncertainty, this kind of monitoring is also necessary to be able to intervene in case the impact is negative, or even a negative Black Swan. The timespan after which innovation resistance is not a factor anymore will vary by target user group. It could be determined experimentally beforehand, but how exactly is beyond the scope of this article.

# 4. How to Keep Stakeholders Happy and Still Innovate

Data-driven design approaches are perfect for achieving incremental improvements (Padua, 2018) as well as for KPI-focused stakeholders who love to see a constant stream of nice, green numbers trickling in. But, as stated above, that endangers innovation, and an unhealthy fixation on metrics can hurt users and the business alike (Laubheimer & Moran, 2021; Speicher, 2021a). However, it is also not a viable solution to run only candidates for Black Swan experiments due to the inherent and extreme uncertainty. That would be hard to sell to management. So, how could we best balance this?

Taleb (2007: ch. XIII) proposes a barbell strategy for financial investments: Invest 85–90% into something extremely safe (e.g., government bonds) and 10–15% into something extremely uncertain, but prone to positive Black Swans (e.g., a venture capital fund). If you lose your 10–15%, your safe investment will cushion that, but if the positive Black Swan *does* happen, you get rich.

That barbell strategy can be directly translated to data-driven design. Pour 85–90% of your resources into "safe" digital designs that are validated and tested by the means described in Section 1.2. Use the remaining 10–15% to run potential Black Swan experiments featuring radical, unvalidated, and unexpected digital designs. This way, you can get the best from both worlds—those nice, green numbers as well as the potential to radically innovate and truly stand out from the crowd.

Depending on the number of experiments a business conducts in relation to the number of customers, this could of course mean that a large share of customers might be often confronted with unusual and at first confusing digital designs. Since, due to the nature of Black Swans, most such experiments will fail, businesses employing the above-described barbell strategy would need to monitor long-term effects on customer behavior. Based on that they could determine whether there might be an upper bound for a feasible "Black Swan exposure." This might lead to less than 10% of experimentation resources being spent on potential Black Swans and is subject to further research.

## 5. Related Work

There is very little existing research on the topic of Black Swans in digital design and online experimentation. During my research for this article, I came across three pieces that touch on the topics I describe.

Azevedo et al. (2020) have developed a framework for optimally distributing scarce experimentation resources for maximum impact, depending on "the thickness of the tail of innovation [or idea] quality" (Azevedo et al., 2020: p. 3). They conclude that businesses with thinner tails should focus on "large-scale, statistically powerful experiments" (Azevedo et al., 2020: p. 3). Businesses with fatter tails should "[run] many small-scale experiments, and [discard] any innovation without outstanding success" (Azevedo et al., 2020: p. 3). They mention potential Black Swans on various occasions, but do not address them beyond online experimentation. Their solution is also oblivious of the fact that experimentation itself can be the cause preventing potential Black Swans (cf. Section 1.3). That being said, their framework might prove valuable for efficiently distributing resources individually *within* the two parts of the barbell strategy proposed above.

On *HackerNoon*, Walls (2018) elaborates on why what he calls "emergent design" or "designing for emergence" has more potential to bring about Black Swans in comparison to other kinds of design (including data-driven and user-centered design). This is one potential answer to how one could "fill" the 10–15% in the barbell strategy with potential Black Swan designs.

In Section 3.2, I have mentioned that further research is necessary on how it could be possible to turn Black Swans gray. According to Taleb (2007: p. 397), Gray Swans are extreme events that can be captured through models while Black Swans are really "unknown unknowns." Aparicio, Bacao, and Oliveira (2014) have investigated this relationship in the context of massive open online courses (MOOCs) and propose that "[d]esigning business models for MOOCs can be a way to domesticate Black Swans and turn MOOCs into Gray Swans" (Aparicio et al., 2014: p. 48).

# 6. Conclusions

In today's industry, data-driven approaches to digital design lead to high rates of successful online experiments and a constant stream of incremental improvements. This strong focus on testing and validation, however, significantly reduces the inherent riskiness and unpredictability of design and experimentation, which acts contrary to the potential for innovation. With his Black Swans, Taleb provides us with a sound theory for how unexpected events can carry a large impact and bring about radical, positive changes. In this article, after verifying that it is actually possible to do so, I have applied Black Swan theory to digital design. Subsequently, I have introduced the notions of *Black Swan designs* and *Black Swan experiments*. To combine the best of both worlds—following Taleb's proposed barbell strategy—in a design organization, 85–90% of the resources should be poured into conventional, data-driven design and the remaining 10–15% into designing for and testing potential Black Swans.

One specific stream of future work is to inquire into how this theory could be operationalized. I intend to investigate this together with my colleagues from our digital design team, particularly in terms of *(a)* ways in which candidates for Black Swan designs might be identified in a real industry setting and *(b)* how such candidates could be selected for Black Swan experiments without violating the criteria. This will prove challenging, as per the theory, Black Swans can inherently not be planned. I intend to report on our experiences in the form of a case study.

Generally, I feel that today's possibilities are too exciting to design the ever-same digital products using the ever-same best practices and testing the same things that all others test based on the same digital design playbooks. Consider how the world wide web looked like in the 1990s. Some would argue websites were really ugly back then, but in reality, there was a beautiful diversity since everyone was wildly trying (almost) anything possible and pushing the boundaries (Buckley, 2021a). There were not as many established standards and best practices yet. Nowadays, a majority of websites—especially in e-commerce—all look more or less the same because we do not embrace Black Swans enough.

I want to close here with two quotes. The first one, by Don Norman (2005): "Great design, I contend, comes from breaking the rules, by ignoring the generally accepted practices [...] This egocentric, vision-directed design results in both great successes and great failures. If you want great rather than good, this is what you must do."

And the second one, by Albert Einstein (Goodreads, 2010): "For an idea that does not first seem insane, there is no hope."

# Acknowledgments

Many thanks to Johanna Jagow and Martin Schmitz for their invaluable input; and to my brother Frederic Speicher, Kevin Bauer, and Tomas Marks for providing feedback on earlier drafts of this article. Your help is much appreciated.

# References

Aparicio, M., Bacao, F., & Oliveira, T. (2014). MOOC's business models: Turning black swans into gray swans. In *Proceedings of the International Conference on Information Systems and Design of Communication* (pp. 45-49). New York, NY: ACM.

Azevedo, E. M., Deng, A., Montiel Olea, J. L., Rao, J., & Weyl, E. G. (2020). A/B testing with fat tails. *Journal of Political Economy, 128*(12), 4614-4672.

Bakshy, E., Eckles, D., & Bernstein, M. S. (2014). Designing and deploying online field experiments. In *Proceedings of the 23rd international conference on World wide web* (pp. 283-292). New York, NY: ACM.

Buckley, M. F. (2021a). Contemporary design has lost its soul. Retrieved July 1, 2022, from https://uxdesign.cc/contemporary-design-has-lost-its-soul-a8a43c00d5aa

Buckley, M. F. (2021b). Modern UX design is killing creativity. Retrieved June 26, 2022, from https://uxdesign.cc/modern-ux-design-is-killing-creativity-f8ba0ff9a989

Georgiev, G. (2018). Analysis of 115 A/B Tests: Average Lift is 4%, Most Lack Statistical Power. Retrieved June 27, 2022, from https://blog.analytics-toolkit.com/2018/analysis-of-115-a-b-tests-average-lift-statistical-power/

Goodreads (2010). A quote by Albert Einstein. Retrieved July 1, 2022, from https://www.goodreads.com/quotes/267305-for-an-idea-that-does-not-first-seem-insane-there

Greenberg, S., & Buxton, B. (2008). Usability evaluation considered harmful (some of the time). In *Proceedings of the SIGCHI conference on Human factors in computing systems* (pp. 111-120). New York, NY: ACM.

Kahneman, D., Knetsch, J. L., & Thaler, R. H. (1991). Anomalies: The endowment effect, loss aversion, and status quo bias. *Journal of Economic perspectives, 5*(1), 193-206.

Knapp, J., Zeratsky, J., & Kowitz, B. (2016). *Sprint: How to solve big problems and test new ideas in just five days*. Simon and Schuster.

Kohavi, R., Deng, A., Frasca, B., Walker, T., Xu, Y., & Pohlmann, N. (2013). Online controlled experiments at large scale. In *Proceedings of the 19th ACM SIGKDD international conference on Knowledge discovery and data mining* (pp. 1168-1176). New York, NY: ACM.

Kohavi, R., Deng, A., Longbotham, R., & Xu, Y. (2014). Seven rules of thumb for web site experimenters. In *Proceedings of the 20th ACM SIGKDD international conference on Knowledge discovery and data mining* (pp. 1857-1866). New York, NY: ACM.

Laubheimer, P., & Moran, K. (2021). Campbell's Law: The Dark Side of Metric Fixation. Retrieved June 27, 2022, from https://www.nngroup.com/articles/campbells-law/

Lisefski, B. (2021). Data-Driven Design Is Killing Our Instincts. Retrieved June 26, 2022, from https://modus.medium.com/data-driven-design-is-killing-our-instincts-d448d141653d

Maeda, J. (2019). *How to speak machine: Computational thinking for the rest of us*. Penguin.

Markoff, J. (2005). *What the dormouse said: How the sixties counterculture shaped the personal computer industry*. Penguin.

Roese, N. J., & Vohs, K. D. (2012). Hindsight bias. *Perspectives on psychological science, 7*(5), 411-426.

Narayanan, A., Mathur, A., Chetty, M., & Kshirsagar, M. (2020). Dark Patterns: Past, Present, and Future: The evolution of tricky user interfaces. *Queue, 18*(2), 67-92.

Norman, D. A. (2005). Human-centered design considered harmful. *interactions, 12*(4), 14-19.

Padua, V. (2018). Disrupt or be disrupted - Experimentation is the key to innovation. Retrieved June 27, 2022, from https://www.itproportal.com/features/disrupt-or-be-disrupted-experimentation-is-the-key-to-innovation/

Patel, N. (2021). How to Find a Winning A/B Testing Hypothesis. Retrieved June 27, 2022, from https://neilpatel.com/blog/winning-ab-testing-hypothesis/

Ries, E. (2011). *The lean startup: How today's entrepreneurs use continuous innovation to create radically successful businesses*. Currency.

Sheppard, B., Sarrazin, H., Kouyoumjian, G., & Dore, F. (2018). The business value of design. *McKinsey Quarterly, 2018*(4), 58-72.

Speicher, M. (2021a). Growth Marketing Considered Harmful. *i-com, 20*(1), 115-119.

Speicher, M. (2021b). KPI-centered design. Retrieved June 26, 2022, from https://uxdesign.cc/kpi-centered-design-8d1f4e231a5

Strategic CFO™ (2015). Black Swan Events. Retrieved July 8, 2022, from https://strategiccfo.com/articles/banking-financing/black-swan-events/

Stroud, C. (2021). 13 experts on quantifying the commercial impact of UX research with A/B tests. Retrieved June 27, 2022, from https://www.userzoom.com/ux-library/quantifying-the-impact-of-ux-research-with-ab-tests/

Stryja, C., Dorner, V., & Riefle, L. (2017). Overcoming Innovation Resistance beyond Status Quo Bias - A Decision Support System Approach (Research-in-Progress). In *Proceedings of the 50th Hawaii International Conference on System Sciences* (pp. 567-576). Retrieved from https://aisel.aisnet.org/hicss-50/cl/hci/6/

Sun, C. (2022). The vanishing designer. Retrieved June 26, 2022, from https://www.doc.cc/articles/the-vanishing-designer

Taleb, N. N. (2007). *Der Schwarze Schwan: Die Macht höchst unwahrscheinlicher Ereignisse [The Black Swan: The Impact of the Highly Improbable]*. Penguin Verlag.

Taleb, N. N. (2020). Statistical consequences of fat tails: Real world preasymptotics, epistemology, and applications. *arXiv preprint arXiv:2001.10488*.

VWO (2022). A/B Testing Guide. Retrieved June 27, 2022, from https://vwo.com/ab-testing/

Walls, A. J. (2018). Emergent Design and Black Swans: Designing A Space For Randomness and Awe. Retrieved July 1, 2022, from https://medium.com/hackernoon/emergent-design-and-black-swans-designing-a-space-for-randomness-and-awe-755c3e3782f5

Xu, Y., Chen, N., Fernandez, A., Sinno, O., & Bhasin, A. (2015). From infrastructure to culture: A/B testing challenges in large scale social networks. In *Proceedings of the 21st ACM SIGKDD International Conference on Knowledge Discovery and Data Mining* (pp. 2227-2236). New York, NY: ACM.